\documentclass[12pt]{article}
\usepackage{amscd,amsmath,amssymb,amstext,amsthm,exscale,latexsym}
\usepackage{graphicx,floatflt}
\newcommand {\al}   {\alpha}       \newcommand {\bt}  {\beta}
       
\newcommand {\dl}   {\delta}       \newcommand {\e }  {\epsilon}

\newcommand {\lm}   {\lambda}      
          
\newcommand {\s }   {\sigma}       
         
\newcommand {\vf }  {\varphi}      
         \newcommand {\om}  {\omega}

\newcommand {\pl}   {\partial}     \newcommand {\nb}  {\nabla}
         
       \renewcommand {\exp}{{\sf\,exp\,}}
\renewcommand {\dim}{{\sf\,dim\,}}

\newcommand   {\const}{{\sf\,const}}     \newcommand   {\diag}{{\sf\,diag\,}}
\newcommand   {\ex}{{\sf\,e}}

\renewcommand {\min}{{\sf\,min\,}}
\newcommand   {\obig}{{\sf O}}
\newcommand {\MC}  {{\mathbb C}}   
\newcommand {\ME}  {{\mathbb E}}   
   \newcommand {\MH}  {{\mathbb H}}
   
   \newcommand {\MP}  {{\mathbb P}}
\newcommand {\MR}  {{\mathbb R}}   
\newcommand {\MS}  {{\mathbb S}}
\newcommand {\MU}  {{\mathbb U}}

   \newcommand {\Sn}  {{\textsc{n}}}

\newcommand {\Gu}  {\mathfrak{u}}   
   \newcommand {\Sd}  {{\textsc{d}}}

\begin{document}
\title     {Adiabatic theorem for finite dimensional quantum mechanical systems}
\author    {M. O. Katanaev
            \thanks{E-mail: katanaev@mi.ras.ru}\\ \\
            \sl Steklov Mathematical Institute,\\
            \sl ul.~Gubkina, 8, Moscow, 119991, Russia}
\date      {\today}
\maketitle
\begin{abstract}
A new simple proof of the adiabatic theorem is given in the finite dimensional
case for nondegenerate as well as degenerate states. The explicitly integrable
two level system is considered as an example. It is demonstrated that the error
estimate given by the adiabatic theorem can not be improved.
\end{abstract}
\section{Introduction}
The adiabatic theorem \cite{BorFoc28} occupies one of the central places in
nonrelativistic quantum mechanics because it allows one to find an approximate
solution of the Schr\"odinger equation for the Hamiltonian which varies slowly in
time. It was proved for the first time for discrete (probably, infinite)
Hamiltonian spectrum with some restrictions on possible energy levels crossings.
The proof for nondegenerate energy levels is given, for example, in
\cite{Messia62}. There are many papers treating the adiabatic theorem; the
corresponding references can be found in \cite{Joye92,Teufel03}. Proofs of the
adiabatic theorem are rather complicated.

In the present paper, a new simple proof of the adiabatic theorem for a finite
dimensional quantum mechanical system is given. At first, we propose the
geometrical interpretation of nonrelativistic quantum mechanics in a finite
dimensional case. We demonstrate that the Hamiltonian of the quantum system
defines the components of the local connection form, and the Schr\"odinger
equation specifies the parallel transport of fibers. The one dimensional
manifold corresponding to time is the base manifold, and the structure group is
the unitary group $\MU(\Sn)$ where $\Sn$ is the dimensionality of the Hilbert
space of a quantum mechanical system. The use of the basis consisting of
eigenvectors of the initial Hamiltonian simplifies the proof and makes it
clearer. The proof is given for nondegenerate as well as degenerate states. To
make the proof maximally simple and to emphasize the most essential feature
we assume that the energy levels do not cross each other.

The comparison with the existing proofs of the adiabatic theorem in a finite
dimensional case \cite{Levi81,ArKoNe02R,Arnold95} is given in conclusion.

In conclusion, we give the example of the two level quantum mechanical system
which is explicitly integrated and demostrates that the estimate given by the
adiabatic theorem cannot be improved.
\section{The adiabatic theorem}
Nontrivial geometric structures, in particular, nontrivial connection on a
principal fiber bundle, often arise when solving equations of mathematical
physics. In the present section, a differential geometric interpretation is
proposed of the Schr\"odinger equation and a new simple proof is given of the
adiabatic theorem \cite{BorFoc28} in the finite dimensional case.

In nonrelativistic quantum mechanics, the state of a system is described by the
vector of the Hilbert space (the wave function) $\psi\in\MH$ which depends on
time and some set of other variables depending on the examined problem. The
evolution of the quantum system in time $t$ is described by the Schr\"odinger
equation \cite{Schrod26A,Schrod26B}
\begin{equation}                                                  \label{escheq}
  i\hbar\frac{\pl\psi}{\pl t}=H\psi,
\end{equation}
where $H$ is the self-adjoint operator acting in the Hilbert space $\MH$ and is
called the Hamiltonian of the system, and $\hbar$ is the Planck constant. We
pose the Cauchy problem for the Schr\"odinger equation with the initial
condition
\begin{equation}                                                  \label{echaus}
  \psi(0)=\psi_0,
\end{equation}
where $\psi_0\in\MH$ is the vector in the Hilbert space normalized by unity.

We further put $\hbar=1$ and denote partial derivative on time by the dot atop,
$\dot\psi=\pl_t\psi$.

Let us assume, for simplicity, that the Hilbert space is a finite dimensional
complex space $\MH=\MC^\Sn$ of complex dimensionality $\dim\MH=\Sn$. We consider
Cauchy problem (\ref{escheq}), (\ref{echaus}) in the general case when the
Hamiltonian of the system depends on time $H=H(t)$. To solve this problem, a
basis in the Hilbert $\MH$ space should be chosen. Surely, a solution of the
problem does not depend on the choice of a basis, and it is chosen for
convenience only. We consider two cases.

Let the basis $e_k\in\MH$, $k=1,\dotsc,\Sn$, be orthonormal and fixed,
$\dot e_k=0$. An arbitrary vector can be decomposed with respect to this basis
$\psi=\psi^k e_k$. Then the Hamiltonian is given by the Hermitian
$\Sn\times\Sn$-matrix $H_l^k$, and the Cauchy problem for the Schr\"odinger
equation acquires the form of a system of ordinary differential equations with
initial conditions
\begin{equation}                                                  \label{echfib}
\begin{split}
  i\dot\psi^k=H_l^k\psi^l,
\\
  \psi^k(0)=\psi_0^k,
\end{split}
\end{equation}
were summation is carried out over repeated indices.

Let us now consider a different orthonormal basis $b_k$ which can depend on time
$b_k=b_k(t)$. Such a basis can be more convenient for solving some problems. The
vector in the Hilbert space $\psi$ can also be decomposed with respect to this
basis $\psi=\psi^{\prime k}b_k$. Then Cauchy problem (\ref{echfib}) looks
differently
\begin{equation}                                                  \label{echneb}
\begin{split}
  i\dot\psi^{\prime k}=H^\prime{}_l^k\psi^{\prime l},
\\
  \psi^{\prime k}(0)=\psi^{\prime k}_0,
\end{split}
\end{equation}
where $H^\prime{}_l^k$ are components of the Hamiltonian with respect to the new
basis calculated below. Two bases are interrelated by the unitary transformation
\begin{equation}                                                  \label{ebatra}
  b_k=S_k^l e_l,~~~~S\in\MU(\Sn),
\end{equation}
depending on time in general, $S=S(t)$. The components of the vector in the
Hilbert space are transformed by the inverse matrix
\begin{equation*}
  \psi^{\prime k}=S^{-1}{}_l^k\psi^l.
\end{equation*}
From here the expression follows for the initial Hilbert space vector
$\psi^{\prime k}_0=S^{-1}{}_l^k(0)\psi^l_0$. Rewriting the Schr\"odinger
equation (\ref{echfib}) in the basis $b_k$, we obtain the Hamiltonian components
with respect to the new basis
\begin{equation}                                                  \label{enewha}
  H'=S^{-1}HS+i\dot S^{-1}S=S^{-1}HS-iS^{-1}\dot S,
\end{equation}
where we have omitted matrix indices for simplicity. We see that the Hamiltonian
components are transformed in the same way as components of the local form of
the $\MU(\Sn)$-connection.

Now we can proceed to geometrical interpretation of nonrelativistic quantum
mechanics. Let time take values on the whole real line, $t\in\MR$. Then we have
the principal fiber bundle
$\MP\big(\MR,\pi,\MU(\Sn)\big)\approx\MR\times\MU(\Sn)$ with the base $\MR$,
typical fiber $\MU(\Sn)$, and projection $\pi:~\MP\rightarrow\MR$
\cite{KobNom6369}. This fiber bundle is trivial because the base is the real
line. The Hamiltonian of the quantum system defines the components of the local
$\MU(\Sn)$-connection form (1-form on $\MR$ with values in the Lie algebra):
\begin{equation*}
  A_t=\lbrace iH_l{}^k\rbrace\in\Gu(\Sn).
\end{equation*}
A vector in the Hilbert space $\psi\in\MH$ is a section of the trivial
associated fiber bundle
$\ME\big(\MR,\pi_\ME,\MH,\MU(\Sn),\MP\big)\approx\MR\times\MH$ with the Hilbert
space $\MH$ being the typical fiber. The Schr\"odinger equation has the form of
equality of the covariant derivative to zero,
\begin{equation*}
  \nb_t\psi=\dot\psi+A_t\psi=0,
\end{equation*}
i.e.\ it defines parallel transport of the vector vector in the Hilbert space.
Under a change of a section, components of the connection transform as they
should
\begin{equation*}
  A'_t=S^{-1}A_tS+S^{-1}\dot S,
\end{equation*}
being components of the local connection form. The curvature of this connection
is zero because the base is one dimensional.

The solution of the Cauchy problem for the Schr\"odinger equation
(\ref{escheq}), (\ref{echaus}) does not depend on the choice of a basis.
Therefore, it is chosen for convenience only. Let a vector in the fixed bases
$e_k$ has the form $\psi^k=U_l^k\psi_0^l$, where the unitary matrix $U_l^k(t)$
defines the evolution operator of a quantum system which, by definition,
satisfies the differential equation
\begin{equation*}
  i\dot U=HU,
\end{equation*}
with the initial condition $U_l^k(0)=\dl_l^k$. It is easily to check that the
evolution operator defines the transformation to such a basis in the Hilbert
space $b_k=U^{-1}{}_k^le_l$ where the Hamiltonian is identically equal to zero,
$H'=0$. Thus the vector in the Hilbert space describing the evolution of a
quantum system in this basis has constant components $\psi_0^k$ defined by the
boundary conditions.

Now we define the adiabatic limit and describe the basis $b_k$ which is used in
the proof of the adiabatic theorem. The adiabatic theorem holds for Hamiltonians
that vary slowly in time. Namely, we suppose that the Hamiltonian is a
sufficiently smooth function on the real parameter $\nu=\e t$, where $\e>0$,
which vary on a finite interval $\nu\in[0,\nu_0]$. Then slow changing of the
Hamiltonian means that parameter $\nu$ varies on a finite value for small $\e$
and large $t$. The adiabatic limit is the double limit in the solution of the
Cauchy problem for the Schr\"odinger equation (\ref{escheq}) and (\ref{echaus})
on the interval $[0,t]$:
\begin{equation}                                                  \label{eadlim}
  \e\to0,~~~~t\to\infty,~~~~\text{under condition}~~\e t=\nu=\const.
\end{equation}
In the analysis of this limit, the time $t$ in the Schr\"odinger equation is
more convenient to be replaced by the parameter $\nu$:
\begin{equation}                                                  \label{eschnu}
  i\e \frac{\pl \psi}{\pl\nu}=H(\nu)\psi.
\end{equation}
In this case, the state vector $\psi(\nu,\e)$ depends also on the parameter
$\e$, and the adiabatic limit corresponds to a simple limit $\e\to0$ for all
values of the parameter $\nu$.

The asymptotic solution of the equation of type (\ref{eschnu}) was constructed
in \cite{VlaVol84C,VlaVol85} in the general case.

To prove the adiabatic theorem we need a special basis depending on time. Let
the initial Hamiltonian $H(\nu)$ of a quantum system be given in a fixed basis
$e_k$. Then there exists a unitary matrix $S(\nu)$ which diagonalizes the
Hamiltonian,
\begin{equation}                                                  \label{ediaog}
  S^{-1}H(\nu)S=H_\Sd(\nu)=\diag\big(E_1(\nu),\dotsc,E_\Sn(\nu)\big),
\end{equation}
where $E_1\le E_2\le\dotsc\le E_\Sn$ are energy eigenvalues of the Hamiltonian
$H$ which are supposed to be ordered. It is well known that columns of the
matrix $S$ are components of eigenvectors of the Hamiltonian $H$. The unitary
matrix $S$ is defined ambiguously, and its arbitrariness is used below.

We allow part of the levels to be degenerate. Denote by $\Upsilon_n$ the set of
indices for which $E_j(\nu)=E_n(\nu)$ when $j\in\Upsilon_n$. Of course, any
index in the set $\Upsilon_n$ can be chosen as $n$. If the energy level $E_n$ is
nondegenerate then the set contains one element: $\Upsilon_n=\lbrace n\rbrace$.
We prove the adiabatic theorem in the case when sets $\Upsilon_n$ for all $n$ do
not change in time, i.e.\ energy levels do not cross.

We suppose that the Hamiltonian $H$, energy levels $E_1,\dotsc,E_\Sn$, and the
transformation matrix $S$ depend sufficiently smooth on $\nu$ on the finite
interval $[0,\nu_0]$.

To prove the adiabatic theorem we need the following statement.
\newline{\bf Lemma.} {\it
There exists the unitary matrix $S$ in Eq.(\ref{ediaog}) such that the condition
\begin{equation}                                                  \label{eadcom}
  \left(S^{-1}\frac{dS}{d\nu}\right)_k^j=0,
  ~~~~\forall k\in\Upsilon_j.
\end{equation}}
holds.
\begin{proof}
Consider two cases. Let the energy level $E_k$ be nondegenerate. Then the
transformation matrix $S$ is defined up to multiplication of each column on
phase factor $S_k^j\mapsto S_k^j\ex^{i\al_k(\nu)}$ for all $j=1,\dotsc,\Sn$.
This is due to the arbitrariness in a phase factor choice for the state vector.
Let the phase factor satisfy the equation
\begin{equation*}
  \frac{d\al_k}{d\nu}=i\sum_{j=1}^\Sn S^{-1}{}_j^k\frac{dS_k^j}{d\nu},
\end{equation*}
where summation over $k$ in the right hand side is absent. It is easily
checked that after the transformation, for any solution of this equation, the
following equality holds:
\begin{equation}                                                  \label{ediael}
  \left(S^{-1}\frac{dS}{d\nu}\right)_k^k=0.
\end{equation}
This can be done for all nondegenerate levels simultaneously.

Assume now that all levels are degenerate, $E_1=\dotsc=E_\Sn$. Then the matrix
$S$ is defined up to the unitary transformation
\begin{equation*}
  S\mapsto SW,~~~~W(\nu)\in\MU(\Sn).
\end{equation*}
Let the matrix $W$ satisfy the equation
\begin{equation*}
  \frac{dW}{d\nu}+S^{-1}\frac{dS}{d\nu}W=0,
\end{equation*}
which always has a solution. Then equality (\ref{eadcom}) is fulfilled
after the transformation for all $j,k$ and any solution.

If only part of the levels is degenerate, then the corresponding unitary
transformation has to be employed only for these levels. Thus equality
(\ref{eadcom}) will be fulfilled for all levels with $E_j=E_k$.
\end{proof}
The proof of the adiabatic theorem is given in orthonormal basis (\ref{ebatra})
where the matrix $S$ is chosen as described in Lemma. This basis consists of
eigenvectors of the initial Hamiltonian $H$:
\begin{equation*}
  Hb_k=E_kb_k,
\end{equation*}
and the Hamiltonian $H(\nu)$ is diagonal in it (see Eq.\ref{ediaog}). We denote
the state vector components in the basis $b_k$ by primes as above,
$\psi=\psi^{\prime k}b_k$. Since the Hamiltonian $H$ in this basis is diagonal,
the squared modulus of the $k$-th state vector component
\begin{equation*}
  |(\psi,b_k)|^2=|\psi^{\prime k}|^2,
\end{equation*}
where parenthesis denote the scalar product in $\MH$, is equal to the
probability to find the quantum system in the state $E_k$ at time moment $t$.

To formulate the theorem, we need the function
\begin{equation*}
  \triangle E_n(\nu)=\underset{j,\s}{\min}|E_j(\s)-E_n(\s)|,~~~~
  \forall\s\in[0,\nu],
\end{equation*}
where minimum $|E_j-E_n|$ is taken over all $j$ for which $E_j\ne E_n$ and all
$\s\in[0,\nu]$. For each value of the parameter $\nu$, the function
$\triangle E_n(\nu)$ is finite because energy levels do not cross each other and
is equal to the minimal distance from the energy level $E_n$ to the remaining
energy levels.
\newline{\bf Adiabatic theorem.} {\it
Let the Hamiltonian $H=H(\nu)$, its eigenstates $b_k(\nu)$, and energy levels
$E_k(\nu)$ be sufficiently smooth functions on $\nu$ on finite interval
$\nu\in[0,\nu_0]$. Suppose that the number of degenerate states is constant in
time. Let $\psi_{(n)}(\nu,\e)$ be the solution of the Schr\"odinger equation
which at the initial moment of time coincides with the eigenstate $b_n(0)$ of
the Hamiltonian $H(0)$ corresponding to the energy level $E_n(0)$. Then
in the adiabatic limit (\ref{eadlim}) the following estimate for the norm holds
\begin{equation}                                                  \label{eotsen}
   1-\sum_{j\in\Upsilon_n}|(\psi_{(n)},b_j)|^2
   =\frac{\obig(\e^2)}{\triangle E^2_n(\nu)},~~~~\forall\nu\in[0,\nu_0].
\end{equation}
That is, the quantum system  during the evolution remains in the eigenstate of
the Hamiltonian $H(\nu)$ corresponding to the energy level $E_n(\nu)$ with
accuracy $\e^2$.}
\begin{proof}
Let us solve the Cauchy problem (\ref{echneb}) in basis (\ref{ebatra}). The
Hamiltonian entering the Schr\"odinger equation in this basis is diagonal up
to linear terms in $\e$,
\begin{equation*}
  H'=H_\Sd-i\e S^{-1}\frac{dS}{d\nu}.
\end{equation*}
Let the matrix $S$ be chosen such as described in Lemma. Suppose that the
system is in the eigenstate of the Hamiltonian $H_\Sd$ at the initial moment of
time and consequently is in the eigenstate of the initial Hamiltonian
$H=SH_\Sd S^{-1}$. This means that the initial condition in the bases $b_k$ has
the form
\begin{equation*}
  \psi_{(n)}(0,\e)=b_n(0)=(\underbrace{0,\dotsc,0}_{n-1},1,0\dotsc,0).
\end{equation*}
Any solution of the Schr\"odinger equation can be written in the form
\begin{equation}                                                  \label{esodia}
  \psi_{(n)}(\nu,\e)
  =\exp\left(-\frac i\e\int_0^\nu \!\!\! d\s H_\Sd(\s)\right)\phi_{(n)}(\nu,\e),
\end{equation}
where $\phi_{(n)}$ is a vector in the Hilbert space $\MH$. Then we obtain the
following equation for the vector $\phi_{(n)}$:
\begin{equation*}
  \frac{\pl\phi_{(n)}}{\pl\nu}=
  -\exp\left(\frac i\e\int_0^\nu \!\!\! d\s H_\Sd\right)
  S^{-1}\frac{dS}{d\nu}
  \exp\left(-\frac i\e\int_0^\nu \!\!\! d\s H_\Sd\right)\phi_{(n)}.
\end{equation*}
We now rewrite the obtained equation with the initial condition in the form of
the integral equation
\begin{equation}                                                  \label{einteq}
  \phi_{(n)}(\nu,\e)=b_n(0)-\int_0^\nu\!\!\! d\s
  \exp\left(\frac i\e\int_0^\s \!\!\! d\lm H_\Sd\right)
  S^{-1}\frac{dS}{d\s}\exp\left(-\frac i\e\int_0^\s \!\!\! d\lm H_\Sd\right)
  \phi_{(n)}.
\end{equation}
For $\e\to0$, the integrand contains fast oscillating factor and can be easily
estimated. Let us consider the modulus of the component of the solution
$\psi^{\prime j}_{(n)}$ corresponding to the eigenstate of the Hamiltonian
$H$ with energy $E_j$ where $E_j\ne E_n$,
\begin{equation}                                                  \label{ecopgj}
  \left|\psi^{\prime j}_{(n)}\right|=\left|\phi^j_{(n)}\right|
  =\left|\sum_{k=1}^\Sn\int_0^{\nu}\!\!\! d\s
  \exp\left(\frac i\e\int_0^\s\!\!\! d\lm(E_j-E_k)\right)
  \left(S^{-1}\frac{dS}{d\nu}\right)_k^j\phi^k_{(n)}\right|.
\end{equation}
The terms with $E_k=E_j$ do note contribute to the sum by virtue of equality
(\ref{eadcom}). For $E_k\ne E_j$, we integrate each term by parts
\begin{multline}                                                  \label{eotska}
  \left.\frac\e{i(E_j-E_k)}
  \exp\left(\frac i\e\int_0^\s\!\!\! d\lm(E_j-E_k)\right)
  \left(S^{-1}\frac{dS}{d\nu}\right)_k^j\phi^k_{(n)}
  \right|_0^\nu-
\\
  -\frac\e i\int_0^{\nu}\!\!\! d\s
  \exp\left(\frac i\e\int_0^\s\!\!\! d\lm(E_j-E_k)\right)
  \frac1{E_j-E_k}\frac d{d\s}\left[
  \left(S^{-1}\frac{dS}{d\nu}\right)_k^j\phi^k_{(n)}\right].
\end{multline}
By assumption, the integrand in the second term is a differentiable function and
can be integrated by parts again. As a result, we obtain that it has the order
of $\e^2$ and can be neglected. The modulus of the first term is evidently
bounded. Thus we obtain the estimate
\begin{equation}                                                  \label{eotsji}
  \left|\psi^{\prime j}_{(n)}(\nu,\e)\right|=
  \frac{\obig(\e)}{\min\big|E_j(\s)-E_k(\s)\big|},~~~~\forall j\notin\Upsilon_n,
\end{equation}
where minimum is taken for all $k$ for which $E_k\ne E_j$, and all
$\s\in[0,\nu]$.

Now we return to expression (\ref{eotska}) again. The function $|\phi_{(n)}^k|$
has the order not less than $\e$ for all $k$ with $E_k\ne E_n$ as the
consequence of estimate (\ref{eotsji}). Therefore contributions of all terms
with indices $k\notin\Upsilon_n$ in sum (\ref{ecopgj}) are no less than $\e^2$
and can be neglected. Hence estimate (\ref{eotsji}) can be improved
\begin{equation*}
  \left|\psi^{\prime j}_{(n)}(\nu,\e)\right|=
  \frac{\obig(\e)}{\min\big|E_j(\s)-E_n(\s)\big|},~~~~\forall j\notin\Upsilon_n.
\end{equation*}
Here minimum is taken only on $\s\in[0,\nu]$.

The norm of any solution is conserved in time and is equal to unity. Thus we
obtain
\begin{equation*}
  1-\sum_{j\in\Upsilon_n}|\psi^{\prime j}_{(n)}(\nu,\e)|^2
  =\sum_{j\notin\Upsilon_n}|\psi^{\prime j}_{(n)}(\nu,\e)|^2,
\end{equation*}
Estimate (\ref{eotsen}) follows from the finiteness of the number of energy
levels.
\end{proof}

In the theorem, the function $\triangle E_n(\nu)$ for each $\nu$ is constant and
can be included in $\obig(\e^2)$. Nevertheless we extracted the factor
$\triangle E_n$ to demonstrate that the assumption that energy levels do not
cross each other is essential. For crossing levels, the denominator in
Eq.(\ref{eotsen}) vanishes, and the proof is not valid.

The adiabatic theorem implies that if a system was initially in the eigenstate
of the Hamiltonian corresponding to the energy level $E_n(0)$ and this level is
nondegenerate, then in the adiabatic limit it will remain in the eigenstate
$E_n(\nu)$ with accuracy of the order of $\e^2$ for finite values of the
parameter $\nu$. If the energy level $E_n$ is degenerate then the system will be
in any of the eigenstates $E_j$ where $j\in\Upsilon_n$ with the same accuracy.
In the next section, we will see that the system can be in any of the degenerate
states $E_j$, $j\in\Upsilon_n$, with probability of the order of unity. Surely,
these statements do not depend on the chosen basis which was used in the proof
of the adiabatic theorem.

Consider now the solution of Cauchy problem (\ref{echfib}) in the adiabatic
limit in the fixed basis in the nondegenerate case. Let $\vf(\nu)$ be the
eigenfunction of the Hamiltonian $H(\nu)$ corresponding to the energy eigenvalue
$E(\nu)$,
\begin{equation*}
  H\vf=E\vf,~~~~\forall \nu\in[0,\nu_0].
\end{equation*}
These eigenfunctions are defined up to a phase factor which can depend on $\nu$.
Let the system be in the eigenstate $\psi_0=\vf(0)$ at the initial moment of
time. In the adiabatic limit, it will be in the eigenstate corresponding to
the energy level $E(\nu)$. The solution of Cauchy problem (\ref{echfib})
can differ from $\vf$ by no more than a phase factor, because the eigenstate is
nondegenerate. Therefore we seek for a solution in the form
$\psi=\ex^{i\Theta}\vf$ where $\Theta(t)$ is unknown function of time. Then
the Schr\"odinger equation yields the equation for the phase
\begin{equation}                                                  \label{eberfa}
  \dot\Theta=i(\dot\vf,\vf)-E.
\end{equation}
The phase is
\begin{equation}                                                  \label{ephaex}
  \Theta(t)=i\int_0^t\!\!ds(\dot\vf,\vf)-\int_0^t\!\!dsE(\e s)
  =i\int_0^\nu\!\!d\s\left(\frac{d\vf}{d\s},\vf\right)-\int_0^t\!\!dsE(\e s),
\end{equation}
because $\Theta(0)=0$ initially.

We now demonstrate that if $\nu\in[0,\infty)$, the phase of the eigenfunction
$\vf$ can be always chosen such that
\begin{equation}                                                  \label{egavfc}
  \left(\frac{d\vf}{d\nu},\vf\right)=0.
\end{equation}
Indeed, let $\vf=\ex^{i\beta}\chi$, where the function $\beta(\nu)$ satisfies
the equation
\begin{equation}                                                  \label{epgasa}
  i\frac{d\beta}{d\nu}=\left(\frac{d\vf}{d\nu},\vf\right)
\end{equation}
with a certain, for example, zero initial condition, $\bt(0)=0$. It is easy
to verify that the equality $(d\chi/d\nu,\chi)=0$ holds for the new
eigenfunctions. Since Eq.(\ref{epgasa}) has always a solution on the half
line, the eigenfunctions $\vf$ of the Hamiltonian can always be chosen in such a
way that equality (\ref{egavfc}) is satisfied.

However, Eq.(\ref{epgasa}) can have no solution on the circle $\MS^1$. Assume
that $\nu\in[0,2\pi]$ on the circle. Then the necessary condition for the
solution existence is the equality
\begin{equation*}
  i\int_0^{2\pi}\!\!\!d\nu\left(\frac{d\vf}{d\nu},\vf\right)
  =2\pi m,~~~~~~m=0,\pm1,\pm2,\dotsc.
\end{equation*}
It is clear that this condition is not fulfilled in the general case. Therefore,
Eq.(\ref{epgasa}) can have no solution on the circle. In this case, the
first term in Eq.(\ref{ephaex}) for the phase can not be eliminated. In essence,
it is the Berry phase.

The solution of the Cauchy problem on the circle $\nu\in\MS^1$ means the
existence of a time machine. These solutions can be rejected as unphysical.
However, Berry proposed another way of reasoning which is considered in the next
paper \cite{Katana11B}.
\section{Two level system}
In this section, we consider the two level quantum mechanical system for which
the Schr\"odinger equation can be solved exactly. We demonstrate that the
estimate given by the adiabatic theorem is unimprovable.

To simplify matters, we shall do the following. We set the diagonal matrix
$H_\Sd$ and the unitary matrix $S$ which define the initial Hamiltonian
$H=SH_\Sd S^{-1}$ instead of specifying the initial Hamiltonian in the fixed
basis and diagonalizing it. Let the diagonal Hamiltonian has the form
\begin{equation*}
  H_\Sd=\begin{pmatrix} E_1(\nu) & 0 \\ 0 & E_2(\nu)\end{pmatrix},
\end{equation*}
where $E_{1,2}(\nu)$ are two given functions. We choose the unitary matrix $S$
in Eq.(\ref{ediaog}) in the form
\begin{equation*}
  S=\begin{pmatrix} \cos\frac\al2 & i\sin\frac\al2 \\[2mm]
  i\sin\frac\al2 & \cos\frac\al2\end{pmatrix},
\end{equation*}
where $\al(\nu)\in\MR$ is also a given function. Consequently, the initial
Hamiltonian is
\begin{equation*}
  H=SH_\Sd S^{-1}=\begin{pmatrix} E_1\cos^2\frac\al2+E_2\sin^2\frac\al2
  & -\frac i2(E_2-E_1)\sin\al \\[2mm] \frac i2(E_2-E_1)\sin\al &
  E_1\sin^2\frac\al2 +E_2\cos^2\frac\al2 \end{pmatrix}
\end{equation*}
and depends on three so far arbitrary functions of the parameter $\nu$.

We solve the Schr\"odinger equation in basis (\ref{ebatra}) in which the
Hamiltonian has the form given by Eq.(\ref{enewha}). Simple calculations yield
the Hamiltonian
\begin{equation*}
  H'=\begin{pmatrix} E_1(\nu) & \frac{\dot\al}2 \\[2mm]
  \frac{\dot\al}2 & E_2(\nu) \end{pmatrix},
\end{equation*}
where the dot denotes differentiation with respect to time $t$. We seek a
solution of Schr\"odinger equation (\ref{echneb}) in the form
\begin{equation*}
  \psi'=\begin{pmatrix}
  \exp\left(-i\int_0^t\!dsE_1\right)\phi \\[2mm]
  \exp\left(-i\int_0^t\!dsE_2\right)\chi \end{pmatrix},
\end{equation*}
where $\phi(t)$ and $\chi(t)$ are two unknown functions. Substitution of this
expression into the Schr\"odinger equation yields the system of equations for
components
\begin{equation}                                                  \label{esypch}
\begin{split}
  i\dot\phi&=\frac{\dot\al}2\exp\left(-i\int_0^t\!\!\!ds(E_2-E_1)\right)\chi,
\\
  i\dot\chi&=\frac{\dot\al}2\exp\left(i\int_0^t\!\!\!ds(E_2-E_1)\right)\phi.
\end{split}
\end{equation}
For $\dot\al\ne0$,
\begin{equation}                                                  \label{echdph}
  \chi=\frac{2i}{\dot\al}\exp\left(i\int_0^t\!\!\!ds(E_2-E_1)\right)\dot\phi.
\end{equation}
as a consequence of the first equation. Differentiating it with respect to time,
we substitute it in the second equation. The result is the second order equation
for $\phi$,
\begin{equation}                                                  \label{eqphis}
  \ddot\phi+\left(i(E_2-E_1)-\frac{\ddot\al}{\dot\al}\right)\dot\phi
  +\left(\frac{\dot\al}2\right)^2\phi=0.
\end{equation}
To solve it explicitly, we specify arbitrary functions entering into the problem
\begin{equation}                                                  \label{earfud}
\begin{split}
  E_1&=E_1^{(0)}+\e t,~~~~E_1^{(0)}=\const,
\\
  E_2&=E_2^{(0)}+\e t,~~~~E_2^{(0)}=\const,
\\
  \al&=2\e t.
\end{split}
\end{equation}
Then equation (\ref{eqphis}) assume the simple form
\begin{equation}                                                  \label{ephedi}
  \ddot\phi+2i\triangle E\dot\phi+\e^2\phi=0,
\end{equation}
where $\triangle E=E_2^{(0)}-E_1^{(0)}$ is the distance between energy levels.
The general solution of this equation depends on two integration constants
$C_{1,2}$:
\begin{equation*}
  \phi=\ex^{-i\triangle Et}\left(C_1\ex^{i\om_\e t}
  +C_2\ex^{-i\om_\e t}\right),
\end{equation*}
where
\begin{equation*}
  \om_\e:=\sqrt{\triangle E^2+\e^2}.
\end{equation*}
The component $\chi$ is given by Eq.(\ref{echdph}). Suppose that initially
the system was in the state $E_1$, i.e.
\begin{equation}                                                  \label{echdat}
  \phi(0)=1,~~~~\chi(0)=0.
\end{equation}
Simple calculations yield the solution of Cauchy problem (\ref{esypch}):
\begin{equation}                                                  \label{esolss}
\begin{split}
  \phi&=\ex^{-i\triangle Et}\left[\cos\left(\om_\e t\right)
  +\frac{i\triangle E}{\om_\e}
  \sin\left(\om_\e t\right)\right],
\\
  \chi&=~\ex^{i\triangle Et}\left[-\frac{i\e}{\om_\e}
  \sin\left(\om_\e t\right)\right].
\end{split}
\end{equation}
We write down also the components of the corresponding eigenstate
\begin{equation}                                                  \label{esolsv}
\begin{split}
  \psi^{\prime 1}&=\ex^{-i\left(\frac{\nu^2}{2\e}
  +E_1^{(0)}\frac\nu\e-\triangle E\frac\nu\e\right)}\left[\cos\frac{\om_\e\nu}\e
  +\frac{i\triangle E}{\om_\e}\sin\frac{\om_\e\nu}\e\right],
\\
  \psi^{\prime 2}&=\ex^{-i\left(\frac{\nu^2}{2\e}
  +E_2^{(0)}\frac\nu\e+\triangle E\frac\nu\e\right)}
  \left[-\frac{i\e}{\om_\e}\sin\frac{\om_\e\nu}\e\right].
\end{split}
\end{equation}
From here it follows that the adiabatic limit for the eigenstate itself does not
exist because its phase goes to infinity. However, the estimate for the
squared modulus of the component can be given. We have the following estimate
for solution (\ref{esolsv})
\begin{equation*}
  1-|\psi^{\prime1}(\nu,\e)|^2=\frac{\obig(\e^2)}{(\triangle E)^2},
  ~~~~|\psi^{\prime2}(\nu,\e)|^2=\frac{\obig(\e^2)}{(\triangle E)^2},
\end{equation*}
which coincides with the estimate in the adiabatic theorem. Hence it follows
that the estimate is unimprovable.

Let us now consider the case of degenerate states $E_1=E_2$ for functions
(\ref{earfud}) specified above. Now Eq.(\ref{ephedi}) is reduced to the
equation of free oscillator:
\begin{equation*}
  \ddot\phi+\e^2\phi=0,
\end{equation*}
and is easily integrated. We write down the solution of the corresponding Cauchy
problem (\ref{echdat}) for the eigenvector components
\begin{equation*}
\begin{split}
  \psi^{\prime 1}&=~~~\ex^{-i\left(\frac{\nu^2}{2\e}-E_1^{(0)}\frac\nu\e\right)}
  \cos\nu,
\\
  \psi^{\prime 2}&=-i\ex^{-i\left(\frac{\nu^2}{2\e}-E_1^{(0)}\frac\nu\e\right)}
  \sin\nu.
\end{split}
\end{equation*}
We see again that the adiabatic limit for the eigenvector does not exist.
However, the squared moduli of the components are well defined
\begin{equation*}
  |\psi^{\prime1}|^2=\cos^2\nu,~~~~|\psi^{\prime2}|^2=\sin^2\nu.
\end{equation*}
As a result, we see that the state vector $\psi'$ oscillates between degenerate
states when the parameter $\nu$ increases. This means that if the system is
initially in one of the degenerate states, then it can be found in any of the
degenerate states with probability of the order of unity in the evolution
process.
\section{Conclusion}
In this work, we give a new simple proof of the adiabatic theorem. To simplify
the proof, we assumed that the Hilbert space is finite dimensional and energy
levels do not cross each other. The transformation to the basis consisting of
eigenvectors of the initial Hamiltonian of the quantum mechanical system (in
which it is diagonal) allowed us to make the proof clearer and to elucidate the
most essential points. Then we considered the example of the two level system
which is solved exactly. It was demonstrated that the probability estimate given
by the adiabatic theorem is unimprovable.

Let us compare the proof presented above with the initial proof. Born and Fock
\cite{BorFoc28} considered the case when the Hamiltonian spectrum was discreet
but can be unbounded. They implicitly made the assumption that energy levels do
not cross each other for almost all time moments. In addition, they accepted
some kind of energy level crossing during the evolution. We considered a simpler
finite dimensional case when energy levels do not cross each other. This allowed
us to simplify the proof and elucidate the most essential features. Estimate
(\ref{eotsen}) is in agreement with the estimate given in \cite{BorFoc28}. Our
proof used the basis in which the initial Hamiltonian is diagonal, and this
allowed us to make the proof clearer. The estimate for integral equation
(\ref{einteq}) is given by Born and Fock in the other way: by series
expansions. Moreover, we admitted the existence of degenerate states for all time
moments in our proof.

Similar proof of the adiabatic theorem for finite dimensional case is given in
\cite{Levi81} for linear Hamiltonian systems. It is known that linear
Hamiltonian systems are described by the Schr\"odinger equation with the special
type Hamiltonian. The idea of the proof is based on such transformation of the
Hamiltonian where the dependence on small parameter is explicit. The proof given
in the present paper can be applied not only to linear Hamiltonian systems but
also to quantum systems of general type. In addition, we used the unitary
transformation for the Hamiltonian instead of symplectic one in \cite{Levi81},
and we think that this simplified the proof.

The proof of the adiabatic theorem for finite dimensional Hamiltonian systems
of general type including nonlinear is given in \cite{ArKoNe02R}. The proof uses
the canonical transformation to the action-angle variables. Surely, it is
applicable for linear systems as well. As it was already mentioned, linear
Hamiltonian systems are equivalent to the particular class of Schr\"odinger
equations and do not include all nonrelativistic quantum systems. In this
respect the proof in the present paper is more general. It includes all finite
dimensional quantum systems and therefore all linear Hamiltonian systems.

The proof of the adiabatic theorem for a finite dimensional quantum mechanical
system for small parameter $\nu$ is given in \cite{Arnold95}. The proof used the
expansion of solutions on parameter $\e$ as well as on $\nu$ and kept only
linear terms. This corresponds to vanishing of the right hand side of the
estimate (\ref{eotsen}) for $\e\to0$ and $\nu\to0$. In the present paper, the
proof of estimate (\ref{eotsen}) is given for small $\e$ and is uniform in
$\nu$ on an arbitrary segment $[0,\nu_0]$. We did not use any expansion.

The author is grateful to I.~V.~Volovich and D.~V.~Treschev for discussions and
fruitful comments. The work is partly supported by the RFBR (grants
11-01-00828-a and 11-01-12114-ofi\_m),  the Program for Supporting Leading
Scientific Schools (Grant No.\ NSh-7675.2010.1), and the program ``Contemporary
Problems in Theoretical Mathematics'' of the Russian Academy of Science.

\end{document}